\documentclass{appolb}
\usepackage{graphicx}

\newcommand{\beq}{\begin{equation}}
\newcommand{\eeq}{\end{equation}}
\newcommand{\ba}{\begin{array}}
\newcommand{\ea}{\end{array}}
\newcommand{\bea}{\begin{eqnarray}}
\newcommand{\eea}{\end{eqnarray}}
\newcommand{\bi}{\begin{itemize}}  
\newcommand{\ei}{\end{itemize}}
\newcommand{\ben}{\begin{enumerate}} 
\newcommand{\een}{\end{enumerate}}
\newcommand{\bc}{\begin{center}}
\newcommand{\ec}{\end{center}}

\newcommand{\De}{\Delta}
\newcommand{\ep}{\varepsilon}

\newcommand{\ptrans}{p_{\rm trans}}
\newcommand{\ntrans}{n_{\rm trans}}
\newcommand{\etrans}{\varepsilon_{\rm trans}}

\newcommand{\destab}{\De\ep_{\rm crit}}

\newcommand{\dd}{\textrm{d}}

\begin{document}
\title{Compact Star Twins with a Dark Matter Core%
\thanks{Presented at the XXVII Cracow EPIPHANY Conference on Recent Advances in Astroparticle Physics (2022).}%
}
\author{David Alvarez-Castillo
\address{Institute of Nuclear Physics, Polish Academy of Sciences, 31-342 Cracow, Poland}
\\[3mm]
{Micha{\l} Marczenko
\address{Incubator of Scientific Excellence - Centre for Simulations of Superdense Fluids,
University of Wroc{\l}aw, plac Maksa Borna 9, PL-50204 Wroc{\l}aw, Poland}
}
}
\maketitle
\begin{abstract}

We present a model of compact stars with a dark matter core. The hadronic equation of state is based on the parity doublet model and does not present a phase transition to quark matter. Instead, a strong first-order phase transition to dark matter described by a constant speed of sound model leads to the scenario of compact star mass twins. Compact star structural properties which obey state-of-the-art measurements and constraints are presented.
\end{abstract}
  
\section{Introduction}

The long-standing problem of the clarification of dark matter (DM) has persisted since its discovery. Astronomical observations, such as anomalies in the rotation curves of galaxies or gravitational lensing, as well as cosmological features, support the idea that DM is present in the universe however hitherto unexplainable. Several attempts to describe DM include exotic particles such as supersymmetric ones or axions, as well as modifications to Einstein's gravity or extra forces beyond the Standard Model. Dark matter is thought to interact only gravitationally and perhaps changing its strength over different scales, as for instance suggested in Chameleon models~\cite{Largani:2019cen}. Consequently, DM might have some sizeable effect on compact stars in whose interiors and surroundings gravity is very strong. Within this work, we focus on neutron stars, motivated by the aforementioned facts. The idea that DM can be found in neutron star (NS) interiors, exteriors, or in both cases~\cite{Guver:2012ba, Ciarcelluti:2010ji} has been explored already in~\cite{RafieiKarkevandi:2021hcc}, where it is meant to be a separated fluid weakly interacting gravitationally with normal matter. Furthermore, examples of microscopic approaches to DM in compact stars that might serve as evidence include axions and their emission resulting in cooling of compacts~\cite{Sedrakian:2015krq}, or sexaquarks~\cite{Shahrbaf:2022upc} which might lead to stellar collapse~\cite{Blaschke:2022knl} into a hybrid star with a quark matter (QM) core.

We consider a particular scenario, where we assume that DM lies in the core of compact stars without inquiring into its nature. We allow the condensed DM to interact with normal matter through nuclear interactions. The resulting dark matter equation of state undergoes a strong first-order phase transition to hadronic matter in the interior of compact stars, producing the so-called \textit{compact star mass twins} configurations, stars of about the same gravitational mass but different radius. This phenomenon has been studied in many works~\cite{Benic:2014jia,Dexheimer:2014pea,Alvarez-Castillo:2017qki,Alvarez-Castillo:2018pve,Sharifi:2021ead,Montana:2018bkb,Cierniak:2021vlf} where compact stars feature a deconfined quark matter core. The astrophysical aspects of the twins have been studied in~\cite{Blaschke:2019tbh}. Moreover, in order to discriminate from the standard mass twins scenario, we choose a hadronic equation of state that does not feature quark deconfinement but includes chiral symmetry restoration. Thus, the resulting hybrid compact stars bear a DM core.

\section{The Equation of State}

The most basic elements of compact star twins require a stiff enough hadronic equation of star capable of describing stable high mass stars of more than the observed 2M$_{\odot}$ pulsars~\cite{Fonseca:2021wxt} and a strong first-order phase transition to high densities. Following the softening at the phase transition, the EoS at the high-density phase should be as well stiff enough to produce stable stars. For compact star twins to appear, the Seidov condition $\De\ep>\destab$~\cite{Seidov:1971} should be fulfilled. The expression
\begin{equation}
\frac{\destab}{\etrans} = \frac{1}{2} + \frac{3}{2}\frac{\ptrans}{\etrans} \ .
\label{eqn:stability}
\end{equation}
relates critical energy density difference at the transition $\destab$ with the energy density and pressure at the onset of DM, $\ptrans$, and $\etrans$.
As for the lower neutron star densities, the NS crust is well described by the approach developed in~\cite{Douchin:2000kad}.

\subsection{Hadronic Equation of State}

Central densities of neutron stars lie up to a few times normal nuclear density, so it is to be expected that baryons change their properties due to the restoration of chiral symmetry. The recent lattice QCD (LQCD) results~\cite{Aarts:2015mma} exhibit a clear manifestation of the parity doubling structure for the low-lying baryons around the chiral crossover at finite temperature. Such properties of the baryonic chiral partners can be described in the parity doublet model~\cite{Detar:1988kn}. It has been applied to hot and dense hadronic matter, neutron stars, as well as the vacuum phenomenology of QCD~(see, e.g.,~\cite{Jido:2001nt, Zschiesche:2006zj, Marczenko:2020omo, Marczenko:2021uaj, Marczenko:2022hyt}).
In the parity doublet model, the masses of the chiral partners, $N_\pm$, are given by
\begin{equation}\label{hadron_mass}
m_\pm = \frac{1}{2} \left(\sqrt{\left(g_1+g_2\right)^2\sigma^2+4m_0^2} \mp \left(g_1-g_2\right)\sigma\right) \textrm,
\end{equation}
where $g_i$'s are baryon-to-meson coupling constants and $m_0$ is the chirally invariant mass parameter. From Eq.~(\ref{hadron_mass}), it is clear that the chiral symmetry breaking generates only the splitting between the two masses. When the symmetry is restored, the masses become degenerate, $m_\pm(\sigma = 0) = m_0$. The positive-parity state, corresponds to the nucleon $N(938)$. Its negative parity partner is identified with $N(1535)$. In this work, we adopt the parametrization for purely nucleonic EoSs in the mean-field approximation~\cite{Marczenko:2022hyt}.

\subsection{Dark Matter Equation of State}

For simplicity, in order to parametrize the microscopically unknown dark matter equation of state we consider the constant speed of sound (CSS) approach, as introduced in~\cite{Alford:2013aca}.
\beq
\ep(p) = \left\{\!
\begin{array}{ll}
\ep_{\rm NM}(p) & p<\ptrans \\
\ep_{\rm NM}(\ptrans)+\De\ep+c_{\rm DM}^{-2} (p-\ptrans) & p>\ptrans
\end{array}
\right.\ ,
\label{EoSqm}
\eeq
where $\ep_{\rm NM}$ is the energy density of nuclear matter, and $c_{\rm DM}$ is the speed of sound of dark matter.

\section{Results}

\begin{figure}
\includegraphics[width=1.0\textwidth]{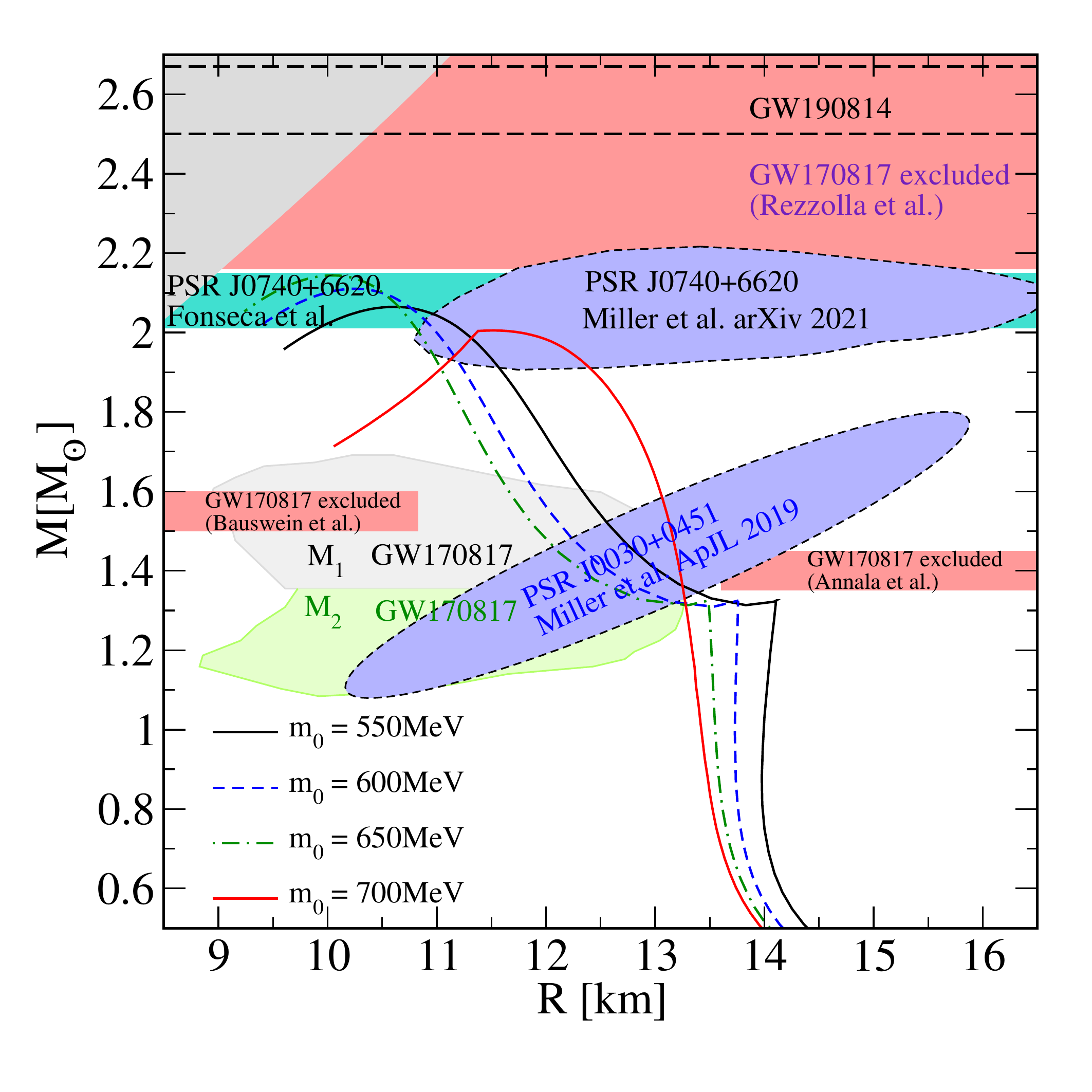}%
\caption{Mass-radius diagram for compact stars which includes measurement regions as well as exclusion regions. Each curve corresponds to a sequence of compact stars for the same parameters within our model which is characterized by the hadronic $m_{0}$ parameter. Description of the several regions in this diagram as well as their usage within Bayesian studies can be found in~\cite{Ayriyan:2021prr}.}
\label{MR}
\end{figure}

\begin{figure*}[!bpht] 
\begin{center}$
\begin{array}{cc}
\includegraphics[width=0.5\textwidth]{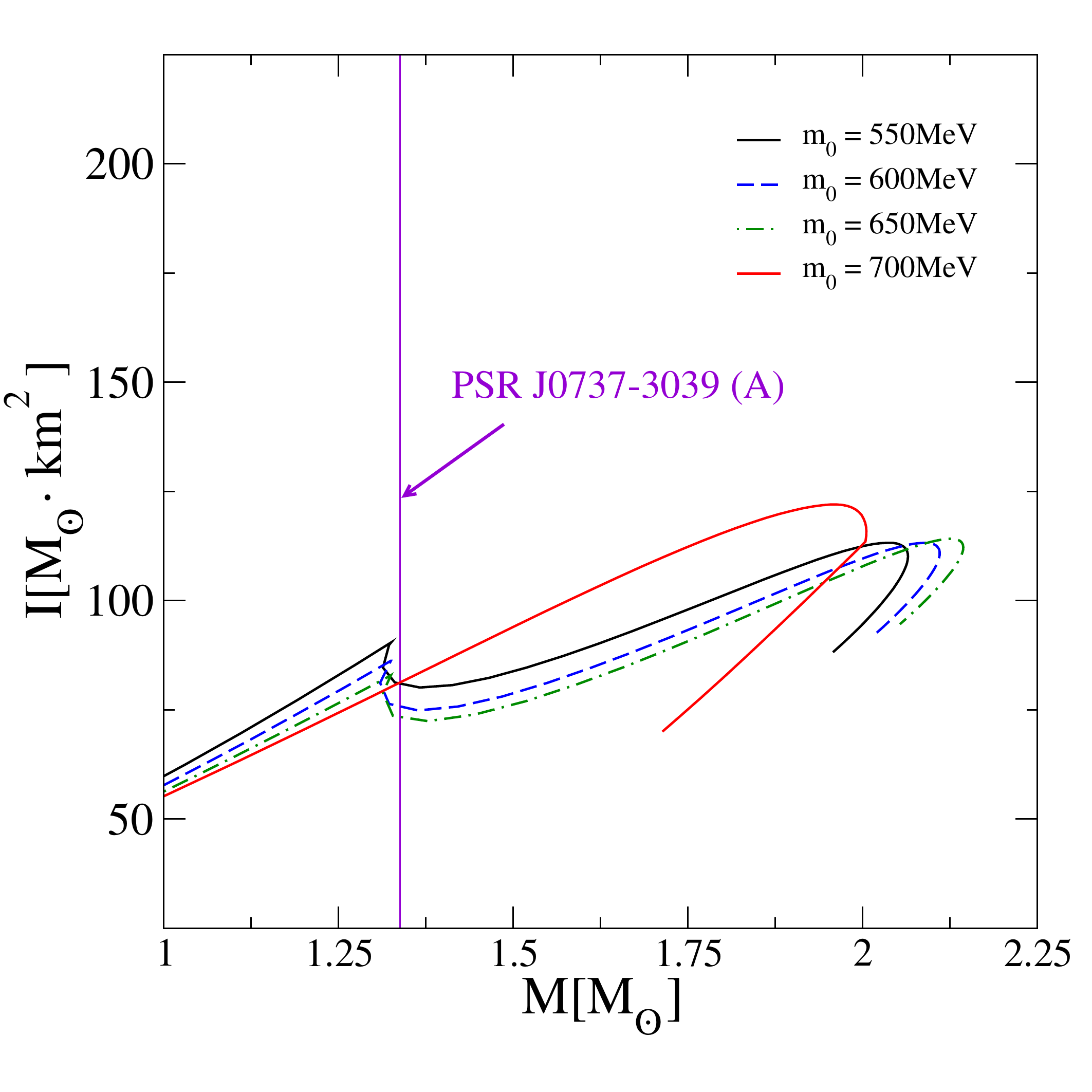} & \hspace{-0.5cm} \includegraphics[width=0.5\textwidth]{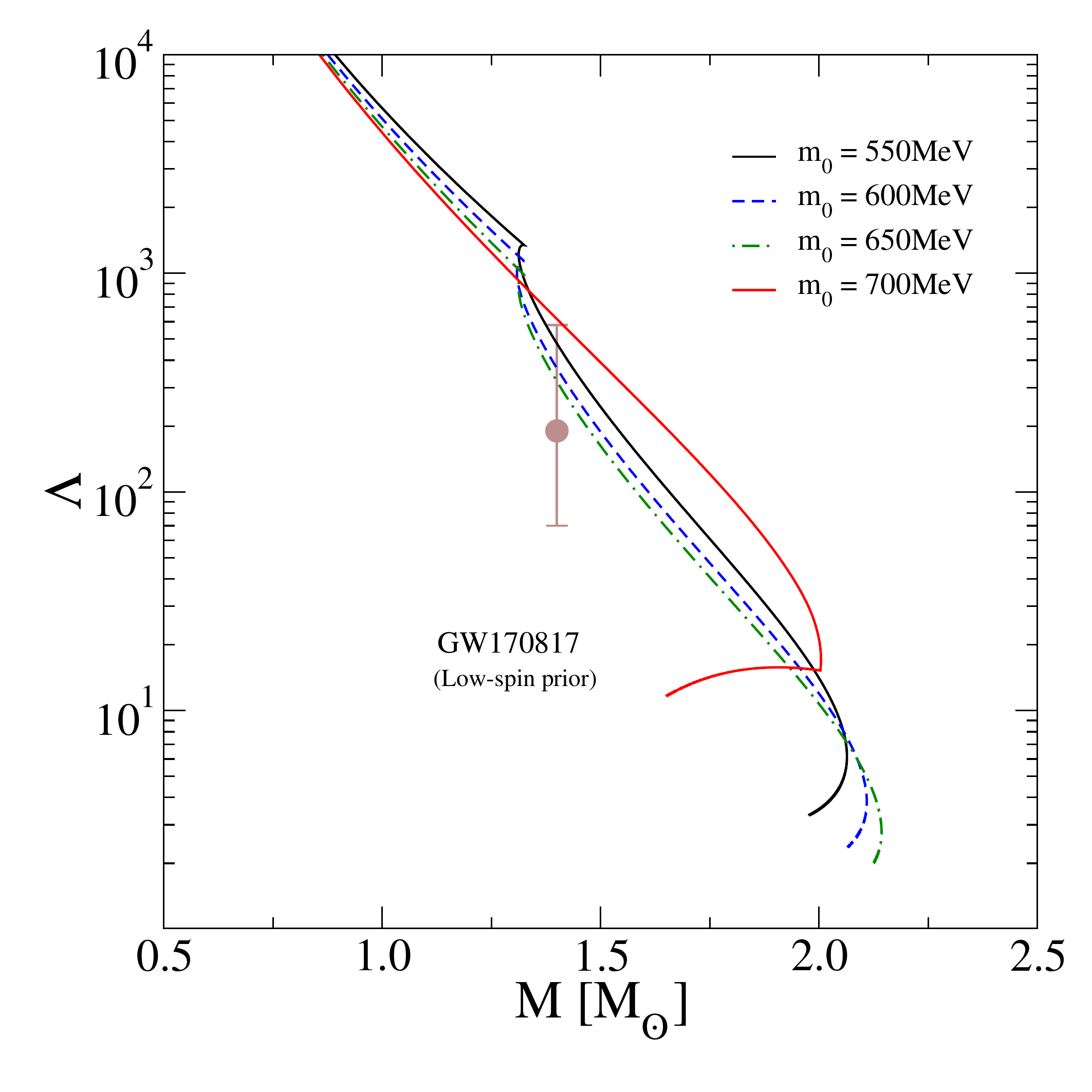}\\
\end{array}$
\end{center} 
\caption{Stellar moment of inertia $I$ (left panel) and dimensionless tidal deformability $\Lambda$ (right panel) related to the compact star mass $M$ with corresponding measurements in electromagnetic and gravitational signals, respectively. A recent Bayesian estimation of the moment of inertia reports $I_{*} = 1.15^{+0.38}_{-0.24} \times 10^{45}$g cm$^{2}$ \cite{Landry:2018jyg}.}
\label{ILM}
\end{figure*}

The resulting sequences of compact stars with dark matter cores are displayed in the mass-radius diagram of Fig.~\ref{MR}. They are obtained by integrating the Tolman-Oppenheimer-Volkoff equations~\cite{Tolman:1939jz, Oppenheimer:1939ne}:
\begin{eqnarray}
 \label{TOV}
\frac{\dd P( r)}{\dd r}&=& 
-\frac{\left(\varepsilon( r)+P( r)\right)
\left(m( r)+ 4\pi r^3 P( r)\right)}{r\left(r- 2m( r)\right)},\\
\frac{\dd m( r)}{\dd r}&=& 4\pi r^2 \varepsilon( r).
\label{eq:TOVb}
 \end{eqnarray}
for each compact star defined by its central density $\varepsilon_c$ up to the maximum mass where $ \partial M /\partial \varepsilon_c> 0$. For the solution of these equations the EoS relation $P(\varepsilon)$ must be used as an input. For our models, It can be seen that the three lower $m_{0}$ EoS parameters can lead to stable configurations at low compact star masses avoiding the excluded region by ~\cite{Annala:2017llu}. The remaining EoS with $m_{0}=700$MeV is soft enough to avoid this constraint and support the 2M$_{\odot}$ compact star, however if dark matter sets in its core it would collapse due to the softness of its nuclear matter possibly leading to a detectable radiation emission event.
In addition, we have computed other stellar properties like the moment of inertia and tidal deformation, which have been already observed through electromagnetic or gravitational radiation of compact stars and their mergers, respectively.
The moment of inertia is computed following \cite{1994ApJ...424..846R}:
\begin{equation}
 I \simeq \frac{J}{1+2GJ/R^{3}c^{2}} ,~~~~~~~~~~J=\frac{8\pi}{3}\int_{0}^{R}r^{4}\left(\rho(r)+\frac{P(r)}{c^2}\right)\Lambda(r)dr.
\end{equation}
with $\Lambda(r)=[1-2Gm(r)/rc^{2}]^{-1}$. Tidal deformability is computed following~\cite{Hinderer:2007mb}. The expression $ \Lambda=2\kappa_{2}R^{5}/3M^{5}$
relates the Love number $\kappa_{2}$ and the mass $M$ and radius $R$ of the star. This quantity has been estimated from the neutron star merger event GW170817~\cite{LIGOScientific:2017vwq,LIGOScientific:2018cki}. Figure~\ref{ILM}
shows the corresponding diagrams for these quantities and astrophysical measurements, whereas table \ref{table_parameters} shows the parameters found for our EoS models.

 %
\begin{table}[h]
\begin{center}
\resizebox{\columnwidth}{!}{%
\begin{tabular}{c|cccccc}
\hline \hline 
Hadronic EoS &	$M_{onset}$ &$\De\ep$& $c_{\rm DM}$ & $\ptrans$ & $\etrans$ & $\ntrans$ \\
$m_{0}$ [MeV] & [M$_{\odot}$] & [$\etrans$] &	[c]  &	[MeV/fm$^{3}$]  & [MeV/fm$^{3}$] & [fm$^{-3}$]        \\
\hline
550 &1.325 &0.75 &0.80 &30.95 & 281.32 & 0.286\\
600 &1.325 &0.75 &0.90 &30.95 & 281.32  & 0.286\\
650 &1.325 &0.70 &0.98 &30.95 & 281.32 & 0.286\\
700 &2.000 &0.75 &1.00 &417.63& 1312.42 & 1.049\\
\hline
\end{tabular}
}
\end{center}
\caption{\label{table_parameters}EoS parameters found in this work.}
\end{table}

\section{Outlook}

The search for DM in NS has become feasible due to the birth of multi-messenger astronomy and progress on the determination of the cold, dense matter EoS. The macroscopical properties of compact stars may be affected from quantities like the cosmological constant~\cite{Largani:2018vus} or the presence of DM, the latter considered here. Observations of the effects of a strong phase transition on supernova explosions, compact stars, and their mergers~\cite{Bauswein:2022vtq} or of NS cooling as well as detection of associated cosmic rays~\cite{CREDO:2020pzy}, eccentric binary orbits~\cite{Alvarez-Castillo:2019apz}, emissions from evolutionary stages~\cite{Alvarez-Castillo:2020nkp}, or the growth of black holes as DM in the interior of rotating NS~\cite{Kouvaris:2013kra}, may allow for testing our NS twins hypothesis.
In our work, we have introduced a realistic equation of state for the nuclear matter which features chiral symmetry restoration without phase transition to deconfined QM in order to produce NS twins with a DM core described by the CSS EoS. Under these specific assumptions, best upon corroboration of the hadronic EoS implemented, detection of the mass twins would allow for the study of the bulk properties of DM. Inclusion of more refined DM models in NS is work in progress.

\section*{Acknowledgements}
The authors thank the COST Action CA16214 ``PHAROS'' for its
support of their networking activities. M.M. acknowledges support from the Polish National Science Centre (NCN) under  No. 2017/27/N/ST2/01973 Preludium Grant, and the program Excellence Initiative–Research University of the University of Wroclaw of the Ministry of Education and Science.

\end{document}